\documentclass[letterpaper,twocolumn]{esapub} 
\usepackage{epsfig}
\title{The effect of turbulent pressure on the p-mode
frequencies in stellar models}
\author{P. Demarque}
\author{L.H. Li}
\author{F.J. Robinson}
\author{S. Sofia}
\affil{Center for Solar \& Space Research, Yale University, New Haven, CT 06520-8101, USA}
\author{Y.-C. Kim}
\affil{Center for Space Astrophysics, Yonsei University, Seodaemoon-gu, Shinchon-dong 134, Seoul 120-749, Korea}
\author{K.L. Chan}
\affil{Mathematics Department, Hong Kong University of Science \& Technology, Hong Kong, China}
\author{D.B. Guenther}
\affil{Department of Astronomy \& Physics, Saint Mary's University, Halifax, Nova Scotia B3A 4R2, Canada}

\begin{document}

\maketitle

\begin{abstract}
We have constructed models for the sun at three stages
of its evolution: a 
zero-age main sequence model,
the present sun, and a subgiant model.  For each model,
the turbulent pressure and turbulent
kinetic energy were 
calculated from 3-d radiative hydrodynamical simulations
(described in the poster by Robinson et al.), and inserted into the
1-d stellar models.  We note that in these simulations,
the turbulent pressure is not a free parameter, but can be computed from the
resulting velocity field.  We show the calculated p-mode frequencies
for the model of the present sun, with and without turbulent pressure, and
compare them to the observed solar frequencies. 
When the turbulent pressure is included in the models, the 
calculated frequencies
are brought closer to the observed frequencies in the sun by up to two 
$\mu Hz$, strictly from structural effects.  The effect of including 
turbulent pressure on p-mode
frequencies is also shown for the zero-age main sequence 
model.  Our models also suggest
that the importance of turbulent pressure increases as the star
evolves into the subgiant region.  We discuss the importance of also 
including
realistic turbulence as well as radiation in the non-adiabatic calculation of
oscillation frequencies.
\end{abstract}

\section{Introduction}
The purpose of this paper is to calculate the effect of including
the turbulent pressure $P_{turb}$, derived from the 3-d radiative
hydrodynamic simulations of Robinson et al. (described in a companion
poster paper), into the hydrostatic structure of
the outer layers of the models, where $P_{turb}$ is known 
to be significant;
and to determine the frequency shift caused by the inclusion 
of $P_{turb}$ on the
p-mode oscillation frequencies.  The turbulent pressure was calculated for
three models of the evolving sun, corresponding to the sun on the
zero-age main sequence (ZAMS), the present sun, and the future
sun as a subgiant.

\section{Evolutionary models for the sun}

An evolutionary track for the sun was constructed which includes pre-main 
sequence evolution and post-main sequence hydrogen burning phases 
on the giant branch.  The Yale Stellar
Evolution code was used.  We also adopted the standard 
assumptions for the calibration of 
the present sun as a standard solar model.    
The input physics included the OPAL opacities and equation of state, 
and   the 
Alexander low temperatures opacities.  Other physical assumptions
were basically as described in Guenther 
\& Demarque 1997).  The evolutionary track is shown in the
theoretical HR-diagram in Fig.~1.

\begin{figure}
\centering
\epsfig{file=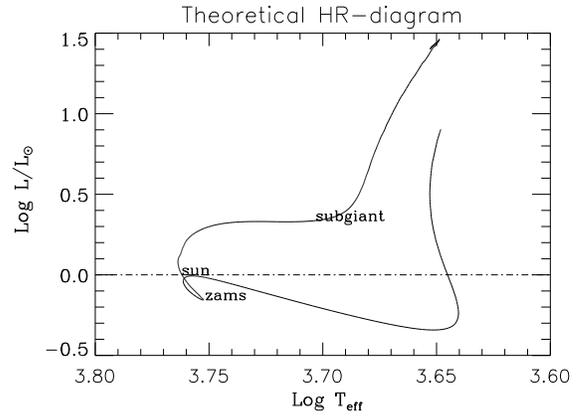, width=8.cm}
\caption{Evolutionary track for the sun from the pre-main sequence stage
to the giant branch
\label{fig1}}
\end{figure}

\section{Radiative hydrodynamical 3-d simulations}
In order to better understand the structure of the atmosphere and of the
highly superadiabatic layer (SAL) along the sequence, a number of
3-d radiative hydrodynamical simulations have been performed using the
hydrostatic 1-d stellar models
as starting points.  The physics (thermodynamics and microscopic physics)
in the simulation was performed in the same way as in the 1-d stellar models.
These simulations follow closely the approach described
by Kim \& Chan (1998), and 
are described in more detail in the companion poster paper
by Robinson et al.  
The full hydrodynamical equations were solved in a thin subsection 
of the stellar model, i.e. a 3-d box located in the vicinity of the 
photosphere.  The radiative transport was treated in the following way:
\begin{itemize}
\item In the deep region of the simulation, where $dS/dz \approx 0$, the diffusion
approximation was used.
\item In the shallow region above, the 3-d Eddington
approximation was used (Unno \& Spiegel 1966).
\end{itemize}  

After the simulation had reached a statistically steady state
(see the poster paper by Robinson et al.), statistical
integration was performed for each simulation, corresponding to 
over 2500 seconds in the case of the solar surface convection.

Some characteristics of the models are illustrated in
Fig.~2 and Fig.~3.
Fig.~2 shows a plot of the ratio of turbulent pressure $\rho w^2$ to the gas
pressure $P$ as a function of $logP$ (i.e. depth),
in the convection simulations for the ZAMS model, the present sun and the
subgiant model, respectively.  We note that the peak in the turbulent pressure
moves outwards, toward lower $logP$, as the sun evolves.

\begin{figure}
\centering
\epsfig{file=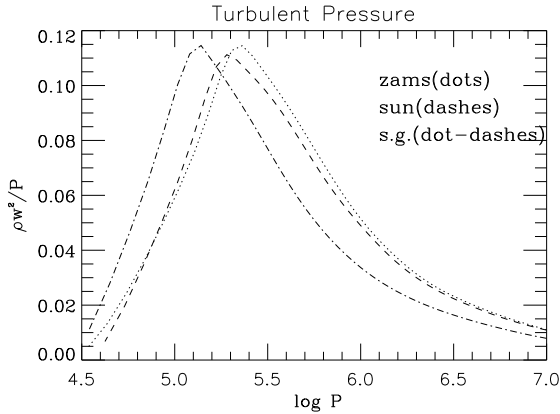, width=8.cm}
\caption{Fraction of turbulent to total pressure in 
the outer layers as a function of depth.
\label{fig2}}
\end{figure}

Fig.~3 compares in addition the kinetic energy density for the two extreme
cases of the ZAMS and
subgiant models, illustrating the effect of evolution on the atmospheric
structure and dynamics.

\begin{figure}
\centering
\epsfig{file=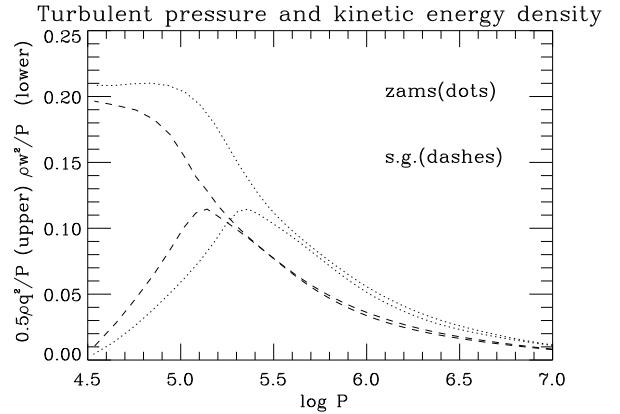, width=8.cm}
\caption{Comparison of the turbulent 
pressure and kinetic energy in the outer layers, 
for the two extreme cases, i.e. 
the ZAMS model and the subgiant model.
\label{fig3}}
\end{figure}

\section{Inserting turbulent pressure into 1-d stellar models}
From the 3-d hydrodynamical simulations, one can derive a turbulent pressure
$P_{turb}$ defined as $P_{turb} = \rho w^{2}$, where $\rho$ is the mean density and
$w$ denotes the turbulent vertical velocity, i.e. the radial direction
in the convection zone. The turbulent energy per unit
mass is $\chi = 0.5 q^{2}$, where $q^{2} = u^{2} + v^{2} + w^{2}$, and the
turbulent velocities $u$, $v$ and $w$ in the $x$, $y$ and $z$ directions
respectively, are
scaled by the sound speed at an arbitrary reference point.

Each parameter, which consists of a mean and a fluctuating part, is computed
from the 3-d statistically averaged flow.  For a given parameter $X$, the
turbulent part is approximated by

$x = \sqrt{ \overline{X^2} - {\overline{X}} ^2 }$, 

where
the overbar denotes horizontal and temporal averaging, and $X$ is the total
quantity (mean plus fluctuating). The turbulent velocities 
$u$, $v$ and $w$ are all computed in this way. 

In the 1-d stellar model, one can define a new parameter $\gamma$ such that
$P_{turb} = (\gamma - 1)\chi\rho$, by analogy with the treatment of magnetic
pressure implemented by Li \& Sofia (2000) in the Yale stellar evolution
code.  The total pressure is thus defined as,
$P_{T} = P + P_{rad} + P_{turb}$, where $P$ and $P_{rad}$ are the
gas and radiation pressures, respectively.

The equation of state is now written as
$\rho = \rho(P_{T},T,\chi,\gamma)$
while the energy conservation equation is modified by the inclusion of
$\chi$, so that

$TdS_{T} = dU_{T} + P_{T}dV - (\gamma - 1)(\chi/V)dV$.

\section{Effect of turbulence on p-mode frequencies}
The effects on the p-mode frequencies
of introducing turbulent pressure in the stellar models is
illustrated in Fig.~4, 5, 6 and 7.  Fig.~4 and Fig.5 refer 
to the model for
the present sun, and Fig.~6 and Fig.~7 to the ZAMS model.
  The oscillation frequencies were
calculated using the non-adiabatic oscillation code developed by Guenther
(1994), which includes non-adiabatic corrections due to the effects of
radiative processes.
The difference between calculated p-mode frequencies and
the observed solar frequencies for non-radial modes with
$\it l$ = 30, 40, 50, 60, 70 ,80, 90 and 100, are plotted vs.
observed frequency.  For each value of $\it l$, the frequency differences for
standard solar models in which the P$_{turb}$ is respectively ignored
(solid lines) and included (dotted lines),
are plotted.  

We see in Fig.~4 and Fig.~5 that
including P$_{turb}$ reduces the model frequencies by approximately
2 $\mu$Hz at the higher frequencies.  Since the effect of
turbulence is only significant near the surface, 
the shift is the largest at
the highest frequencies.  The frequency shift is in the sense of
improving agreement between calculated and observed solar frequencies
(Antia \& Basu 1997; Demarque et al. 1999).

\begin{figure}
\centering
\epsfig{file=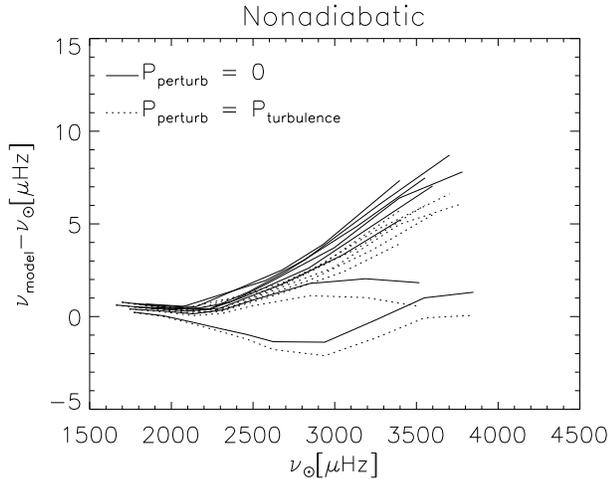, width=8.cm}
\caption{Difference between the present sun model and observed solar
p-mode frequencies vs. observed frequency. Non-adiabatic frequencies are
shown in this plot,
for models with (dotted lines) $P_{turb}$ and without (solid lines),
for different $\it l$ values.
\label{fig4}}
\end{figure}

\begin{figure}
\centering
\epsfig{file=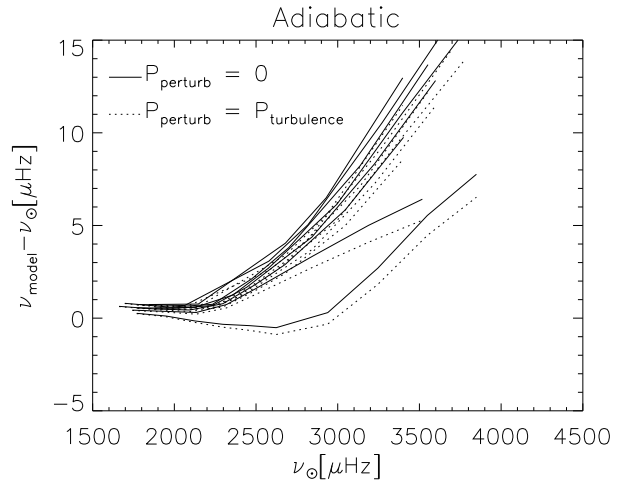, width=8.cm}
\caption{Same as Fig.~4, using adiabatic frequencies.
\label{fig5}}
\end{figure}

We point out that Fig.~4 and Fig.~5 show the p-mode frequency shifts due to
structural changes in the stellar models, but do not include frequency shifts 
which might be due to turbulence in the calculation of the oscillation frequencies.
As mentioned above, while radiation is included in 
Guenther's non-adiabatic
frequency calculations, no account is taken of the Reynolds stress
due to turbulence.

Balmforth (1992) has investigated the effect of turbulence,
and found it to be relatively
larger than the structural effect on the calculated frequencies, of the
order of $10\mu Hz$.  Guzik \& Swenson (1997) found a shift of a
few $\mu Hz$ only for low degree mode ($\it l$ less than 10), and an
increase of up to a few tens $\mu Hz$ for $\it l\geq 100)$.
More recent studies of the Reynolds stress
correction on the
p-mode frequencies have yielded smaller frequency shifts both using a
simple mixing length model (B\"{o}hmer \& R\"{u}diger 1999), 
and a model of
convection which includes a more realistic turbulence energy spectrum
(Bi \& Xu 2000).  An obvious next step is to consider the
combined non-adiabatic effects of radiation and turbulence using 
the turbulent pressure derived from a 3-d physically realistic 
simulation. 

\begin{figure}
\centering
\epsfig{file=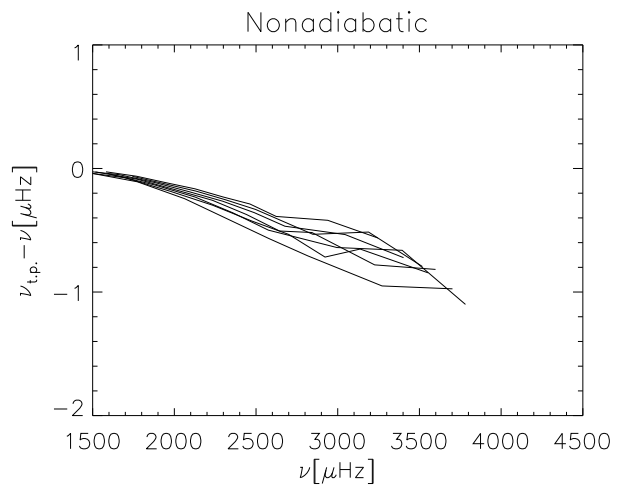, width=8.cm}
\caption{Frequency differences between ZAMS models with 
and without  $P_{turb}$,  
plotted vs. frequency. Non-adiabatic 
frequencies have been used,
for different $\it l$ values.\label{fig6}}
\end{figure}

\begin{figure}
\centering
\epsfig{file=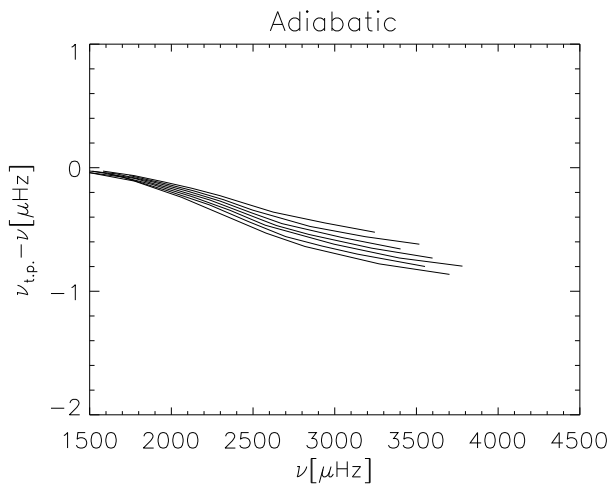, width=8.cm}
\caption{Same as Fig.6, but for adiabatic p-mode frequencies.\label{fig7}}
\end{figure}

Fig.~6 and Fig.~7 show the difference between the calculated p-mode
frequencies for ZAMS models of the sun, with and without the
effects of the turbulent pressure. 
We note that the effect of including
the turbulent pressure in the ZAMS model is smaller than in the present
sun model.

The simulation for the subgiant model indicate that the role of turbulent
pressure increases as the star evolves toward the giant branch.  This
result is likely to be due to densities in the vicinity of the
highly superadiabatic layer that are lower in giants than in main
sequence stars.

Work is now in progress to extend this simulation grid, and to use the
simulations to derive more realistic surface boundary conditions 
along evolutionary tracks than 
are currently provided by the mixing length theory of convection.
   
\section*{Acknowledgements}  This research was supported in part by NASA grant
NAG5-8406 to Yale University.  Support from the
Creative Research Initiative Program of The Korean Ministry of
Science and Technology (Y.-C. Kim) and from the National Science and
Engineering Research Council of Canada (D. B. Guenther) are also
gratefully acknowledged.

\end{document}